\documentclass[english,prb, twocolumn, superscriptaddress]{revtex4}
\usepackage[T1]{fontenc}
\usepackage[latin9]{inputenc}
\usepackage{color}
\usepackage{graphicx}

\makeatletter
\@ifundefined{textcolor}{}
{%
 \definecolor{BLACK}{gray}{0}
 \definecolor{WHITE}{gray}{1}
 \definecolor{RED}{rgb}{1,0,0}
 \definecolor{GREEN}{rgb}{0,1,0}
 \definecolor{BLUE}{rgb}{0,0,1}
 \definecolor{CYAN}{cmyk}{1,0,0,0}
 \definecolor{MAGENTA}{cmyk}{0,1,0,0}
 \definecolor{YELLOW}{cmyk}{0,0,1,0}
}

\makeatother

\usepackage{babel}
\begin{document}

\title{Self-organized pattern formation in laser-induced multiphoton ionization}

\author{Robert Buschlinger }

\affiliation{Institute of Optics, Information and Photonics, University of Erlangen-N�rnberg,
91058 Erlangen, Germany}

\author{Stefan Nolte }

\affiliation{Institute of Applied Physics, Abbe Center of Photonics, Friedrich-Schiller-Universit�t
Jena, 07743 Jena, Germany }

\affiliation{Fraunhofer Institute for Applied Optics and Precision Engineering,
07745 Jena, Germany}

\author{Ulf Peschel }

\affiliation{Institute of Optics, Information and Photonics, University of Erlangen-N�rnberg,
91058 Erlangen, Germany}
\begin{abstract}
We use finite-difference time-domain modelling to investigate plasma
generation induced by multi-photon absorption of intense laser light
in dielectrics with tiny inhomogenities. Plasma generation is found
to be strongly amplified around nanometer-sized inhomogeneities as
present in glasses. Each inhomogeneity acts as the seed of a plasma
structure growing against the direction of light propagation. Plasma
structures originating from randomly distributed inhomogeneities are
found to interact strongly and to organize in regularly spaced planes
oriented perpendicularly to the laser polarization. We discuss similarities
between our results and nanogratings in fused silica written by laser
beams with spatially homogeneous as well as radial and azimuthal polarization.
\end{abstract}


\maketitle
Many dielectrics as e.g. silica glasses are known to be transparent
within a wide frequency range. Only at high intensities absorption
becomes possible, as electrons are promoted to the conduction band
by nonlinear ionization processes.\cite{0034-4885-54-10-002} The
strong intensity dependence of multiphoton ionization allows for the
selective excitation and laser-induced modification of a small focal
region situated inside a material volume. Different kinds of material
modification have been observed, including refractive index changes,\cite{Davis:96}
void formation\cite{glezer:882} and subwavelength volume grating
formation.\cite{eps20999,PhysRevLett.96.057404,Hnatovsky2011} 

Previous modelling efforts concerning laser energy deposition in dielectrics
have concentrated on the temporal and spatial evolution of the laser
pulse itself, while treating the material as homogeneous.\cite{0953-4075-41-2-025601,Bourgeade2010,raey,mezel:093504}
As far as nonlinear self-organization is concerned, a certain seed
is required to start the process. Therefore, we follow a different
approach and investigate the interaction of laser light with nanometer-sized
inhomogeneities. This is of fundamental interest due to the inherent
inhomogeneity of amorphous materials like silica glasses.\cite{JACE:JACE461}
Such inhomogeneities have also been suggested to play a major role
in volume nanograting formation.\cite{0953-4075-40-11-S03}

Our simulations are based on the standard parameters which can be
found in the literature. A good overview of the parameters of laser
light and free carriers present during nanograting formation has been
given by Bulgakova et al.\cite{raey}. There, the intensities achieved
by focussing and nonlinear propagation inside the homogeneous material
cause smooth, submetallic carrier density distributions. We use similiar
parameters, but in our case material inhomogeneities increase the
local intensity and cause the formation of plasma spots.\emph{ }We
demonstrate, that the ionization process is\emph{ }independent of
the exact shape and nature of the initial inhomogeneities. However
we can identify two regimes depending on the local carrier densities
that are reached during irradiaton. For low carrier densities, ionization
enhancement remains confined to the initial region of field enhancement
close to an inhomogeneity. For higher carrier densities, a single
nanoplasma forming at an inhomogeneity site enhances further ionization
in its vicinity and acts as a seed for the growth of an extended structure.
This regime of strong interaction is similar to the ionization instability
which has been predicted for tunneling ionization in gasses\cite{PhysRevE.84.036408,PhysRevLett.102.015002}
and which has been suggested to play a similiar role in the ionization
of transparent dielectrics.\cite{raey} In both cases, intricate structures
with high carrier densities can be formed. However, the local field
enhancement around an existing nanoplasma plays a much more important
role in our case.

We further examine optical self-organization in material systems with
randomly distributed nanometer-sized inhomogeneities. We observe the
formation of planar structures aligned perpendicularly to the laser
polarization with a self-organized period related to the laser wavelength\emph{.
}We discuss the similarities to experimentally observed nanograting
damage patterns and the possible connection of our results to these
phenomena.

\section{Numerical Model\label{sec:Numerical-Model}}

For our model, we use a nonlinear finite-difference time-domain (FDTD)
approach, which previously has been applied to the modelling of ionization
and void formation in silica.\cite{Penano2005,mezel:093504} Maxwells
equations
\[
\frac{\partial}{\partial t}\vec{D}=\frac{1}{\mu_{0}}\nabla\times\vec{B}-\vec{J}
\]

\begin{equation}
\frac{\partial}{\partial t}\vec{B}=-\nabla\times\vec{E}
\end{equation}
 with $\vec{D}=\varepsilon\vec{E}+\vec{P}$ are solved using the standard
FDTD algorithm.\cite{Yee66numericalsolution,taflove:2005} The response
of the unexcited medium, which is dominated by valence band electrons,
is included in the background permittivity $\varepsilon=n^{2}\varepsilon_{0}$,
using a linear refractive index $n=1.45$. 

We also incorporate the Kerr effect using the third-order material
polarization $\vec{P}=\varepsilon_{0}\chi_{3}E^{2}\vec{E}$. This
formulation assumes a scalar third order susceptibility $\chi_{3}=2\times10^{-22}m^{2}V^{-2}$,\cite{PhysRevB.73.214101}
which is a good approximation for linearly polarized light propagating
in glass. 

The remaining contributions to the material response are included
via the current density $\vec{J}=\vec{J}_{d}+\vec{J}_{mpi}$, where
the ionization current $\vec{J}_{mpi}$ is used to model the energy
loss of the electric field due to multiphoton ionization, which excites
electrons to the conduction band. $\vec{J}_{d}$ describes the optical
response of these newly generated conduction band electrons based
on a Drude model
\begin{equation}
\frac{\partial}{\partial t}\overrightarrow{J_{d}}=-\nu_{e}\overrightarrow{J_{d}}+\frac{e^{2}}{m_{e}}\rho\overrightarrow{E}.\label{eq:Jdrude}
\end{equation}
\emph{ }In a complete model, the electron collision frequency $\nu_{e}$
would have to be assumed to depend on carrier density and temperature.
As a simplification, we assume it to have a constant value $\nu_{e}=10^{14}s^{-1}$
lying in the range of reported values.\cite{PhysRevLett.89.186601,mao:697,raey}
As long as the resonant field-enhancement close to a nanoplasma is
not completely damped by collisions, final results have turned out
to be mostly independent of $\nu_{e}$.

The time dependent conduction band carrier density $\rho$ is described
with a rate equation taking into account multiphoton ionization and
recombination 
\begin{equation}
\frac{\partial}{\partial t}\rho=\left(\rho_{0}-\rho\right)\nu_{mpi}-\frac{\rho}{\tau_{rec}}.\label{eq:CarrierDensity}
\end{equation}
In this model, the free carrier density reaches saturation at a value
of $\rho_{0}=2\times10^{28}m^{-3}$.\cite{Penano2005} The electron
recombination time is $\tau_{rec}=150\times10^{-15}s$.\cite{PhysRevLett.73.1990}
For an excitation wavelength of $\lambda=800nm$ and a fused silica
target with a band gap of $W_{ion}=9eV$, $6$ photons are needed
to promote an electron to the conduction band, resulting in an ionization
rate 
\begin{equation}
\nu_{mpi}=\frac{\sigma_{6}I^{6}}{\rho_{0}}
\end{equation}

with a cross-section of $\sigma_{6}=2\times10^{-65}m^{9}W^{-6}s^{-1}$.\cite{PhysRevB.73.214101}
An expression for $\vec{J}_{mpi}$ can be derived by equating the
energy gain of electrons $\frac{\partial}{\partial t}W=W_{ion}\nu_{mpi}\left(\rho_{0}-\rho\right)$
due to multiphoton ionization to the energy loss of the electric field
$\vec{J}_{mpi}\vec{E}$, yielding 
\begin{equation}
\overrightarrow{J}_{mpi}=\frac{\sigma_{6}}{\rho_{0}}W_{ion}I^{5}\overrightarrow{E}\left(\rho_{0}-\rho\right).\label{eq:Jmpi}
\end{equation}

The nonlinear equations that describe the electric field, ionization
loss and carrier density are solved using a fixed-point iteration
method at each FDTD-timestep. 

Additional attention has to be paid to the modelling of inhomogeneities.
As glass is an amorphous solid, nanosize inhomogeneities are always
present due to local variations in the chemical structure\cite{Richter:12}
and to actual voids or gas inclusions.\cite{JACE:JACE461,Hasegawa2000}\emph{
}It has also been reported, that material nonlinearities can be enhanced
by a history of previous laser-irradiation.\cite{Richter:12,PhysRevLett.97.253001,Lancry:11}
Such effects could lead to an additional inhomogeneity in the nonlinear
response. According to literature, voids in conventional silica take
up a fraction of $3\%$ of the material volume\cite{JACE:JACE461}
and have an average size of $d=0.6nm$,\cite{Hasegawa2000}\emph{
}resulting in a mean spacing below $20nm$. \textcolor{black}{Such
voids should be an appropriate model for a typical inhomogeneity.
}As will be shown in section \ref{sec:Simulation-Results}, ionization
around a void leads to the same final plasma structure as ionization
around a region with an enhanced ionization cross-section. Additionally,
final results do not depend on the actual shape of the inhomogeneity,
as long as the initial size does not exceed a few nanometers. This
is to be expected, since scattering from small objects is dominated
by the dipole mode and does not depend on the specific shape.\cite{Bohren_Huffman_1983}
This means that once initial results are established with a fine discretization
($\triangle x=0.5nm$) and realistic inhomogeneities, we can safely
use comparatively large seed inhomogeneities and coarse discretizations.
Nevertheless, to properly resolve the inhomogeneities and the induced
plasma structures, the resolution should not exceed $\triangle x=5nm$.
As the simulation of multiple laser pulses and the subsequent material
modifications are outside the reach of present computational resources,
we limit our scope to continous illumination in order to understand
the potential of purely optical self-organization processes. This
approach can be viewed both as a limiting case of single-pulse excitation
and as an extrapolation of the excitation with multiple pulses.

In all presented cases, the simulation volume is situated deep inside
the bulk of the material. At the material boundaries far away from
the focal region of the laser, intensities are considerably lower.
We will start out with simple geometries where the exciting field
is approximated as a plane wave and then proceed to more realistic
simulations with focussed sources. The simulation volume is terminated
by perfectly matched layers\cite{taflove:2005} in propagation direction.
In transverse direction, we choose periodic boundary conditions. Initially,
the material is taken to be unexcited and the conduction band carrier
density is set to $\rho=0$.

\section{Simulation Results\label{sec:Simulation-Results}}

\begin{figure}
\includegraphics[scale=0.55]{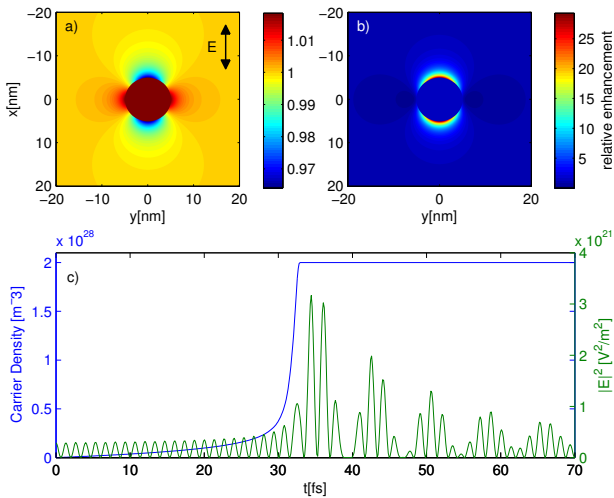}\caption{\label{fig:Local-intensity-enhancement}(color online). Scattering
on a small inhomogeneity. a) and b): Local intensity enhancement for
plasma spheres with different carrier densities $\rho$ (to illustrate
the field structure, calculations are done using the electrostatic
approximation. Refractive indices are obtained from the Drude model).
(a) $\rho=10^{26}m^{-3}\ll\rho_{Mie}$. b) $\rho=2\times10^{28}m^{-3}>\rho_{Mie}$).
c): Temporal evolution of the carrier density and electric field in
a small ionizable sphere in glass irradiated with a plane wave (amplitude
$E_{0}=1.7\times10^{10}Vm^{-1}$). }
\end{figure}
Our simulations focus on the interaction of intense laser-light with
materials with randomly distributed nanometer-sized inhomogeneities.
To get a feeling for the basic processes, we start with a very simplified,
but quasi-analytical model. We consider a \textcolor{black}{spherical
subwavelength inhomogeneity, in which plasma generation can occur
due to a non-vanishing ionization cross-section. Plasma generated
inside the sphere according to Eq. \ref{eq:CarrierDensity} causes
a decrease of the dielectric constant according to Eq. \ref{eq:Jdrude}.
In the case of a sphere much smaller than the exciting wavelength,
the fields can be calculated in a quasi-static approximation}\cite{Bohren_Huffman_1983,jackson_classical_1999}\textcolor{black}{{}
(Figs. \ref{fig:Local-intensity-enhancement}(a),(b)).} Combined with
envelope approximations of Eqs. \textcolor{black}{\ref{eq:Jdrude}}
and \textcolor{black}{\ref{eq:CarrierDensity}}, this model can also
serve to estimate the temporal evolution of the carrier density and
electric field strength inside the sphere (Fig. \ref{fig:Local-intensity-enhancement}(c)).
According to the electrostatic approximation, a dipole wave is excited
at the inhomogeneity site and interferes with the incident plane wave.
Intensity is enhanced both inside the sphere and at its equator perpendicular
to the incident electric field vector (Fig. \ref{fig:Local-intensity-enhancement}(a)).
The free electron density inside the sphere increases due to the positive
feedback between the local electric field and the plasma refractive
index (Fig. \ref{fig:Local-intensity-enhancement}(c)). Eventually,
the plasma reaches a carrier density where the dipole-resonance of
the sphere comes close to the excitation frequency ($\varepsilon_{plasma}(\rho_{Mie},\omega_{source})=-2\varepsilon_{background}$).
\begin{figure}
\includegraphics[scale=0.55]{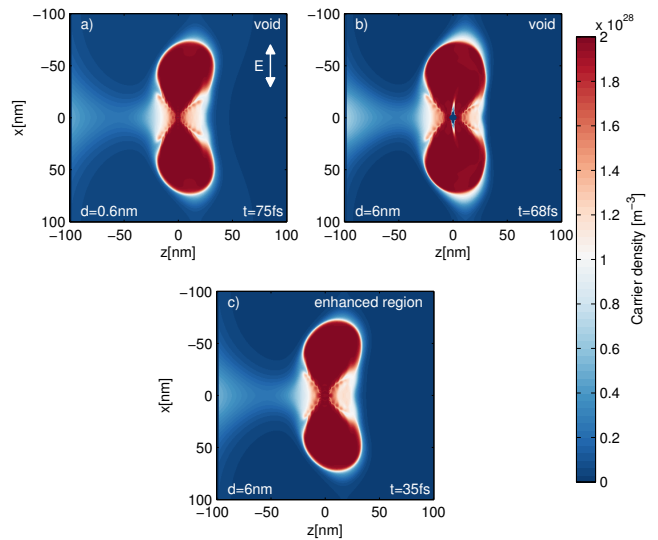}\caption{\label{fig:Early-stages-of}(color online). Carrier density (2D FDTD)
for early stages of ionization around spherical voids ($\sigma_{6,inh}=0$,
$n=1.0$) with different sizes as indicated in the figure as well
as a region with an enhanced ionization cross-section $\sigma_{6,inh}=\sigma_{6}\times5$.
Illumination is a continous plane wave with amplitude $E_{0}=1.7\times10^{10}Vm^{-1}$
normally incident from the left in all three cases.}
\end{figure}
At this point, carrier densities increase almost exponentially. The
scattered field is now strongly enhanced and solely determines the
nearfield intensity pattern. Due to the strong field enhancement,
ionization continues even up to the maximum carrier density $\rho=\rho_{0}$,
where saturation sets in. Pronounced intensity maxima now lie at the
poles of the nanoplasma (Fig.  \ref{fig:Local-intensity-enhancement}(b)). 

If ionization in the surrounding medium is taken into account, this
field enhancement leads to the formation of an ionized region growing
into the direction of the electric field. Since scattering is now
mostly caused by the induced nanoplasma, the final structure is invariant
concerning the microscopic details like shape, size and chemical nature
of the seed inhomogeneity. Even if the initial inhomogeneity is a
void instead of a region with an enhanced nonlinearity, field enhancement
around the void leads to an amplified plasma generation, resulting
in a similiar evolution if a certain threshold is exceeded. To verify
these guesses, we turn our attention to the complete FDTD-based model
as described in section \ref{sec:Numerical-Model}. 

We initially restrict simulations to a two-dimensional geometry, since
this has shown to illustrate the growth process more clearly than
the full three-dimensional case. For our later simulations we will
return to three-dimensional geometries and show that the plasma structures
growing in random media tend to reproduce the features seen in a 2D
simulation. 

\begin{figure}
\includegraphics[scale=0.55]{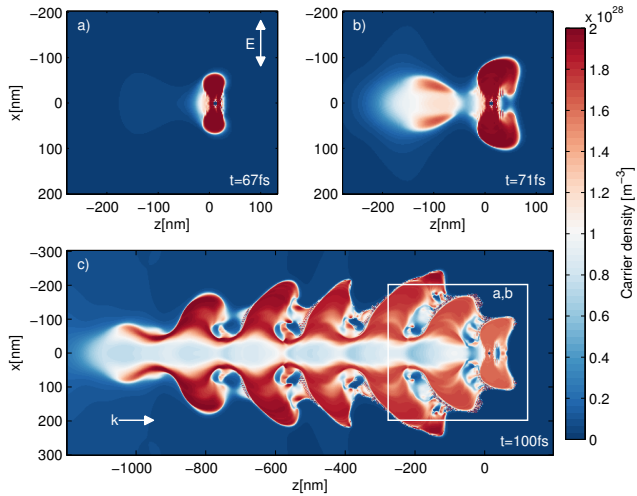}\caption{\label{fig:Plasma-density-2D}(color online). Plasma density (2D FDTD)
around a single void ($\sigma_{6,inh}=0$, $n=1.0$, $d=7nm$) at
the coordinate origin for several stages of structure growth. a) Plasma
growth into the polarization direction. b) Saturation of growth and
initiation of a second structure. c) Periodic plasma structure formed
by subsequent growth. Illumination is a cw plane wave with amplitude
$E_{0}=1.7\times10^{10}Vm^{-1}$ incident from the left. }
\end{figure}

\begin{figure}
\includegraphics[scale=0.55]{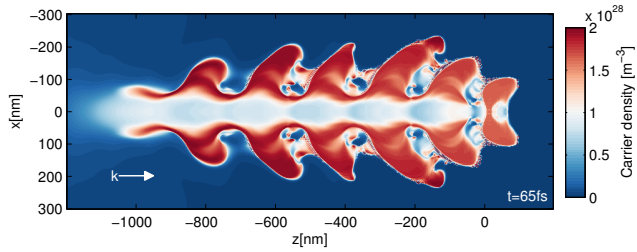}

\caption{\label{fig:EnhancedInhom}(color online). Plasma density (2D FDTD)
around a single inhomogeneity with an enhanced ionization cross-section
($\sigma_{6,inh}=\sigma_{6}\times15$, $d=7nm$) at the coordinate
origin. Illumination is a cw plane wave with amplitude $E_{0}=1.7\times10^{10}Vm^{-1}$
incident from the left. }
\end{figure}

We consider regions with an enhanced ionization cross-section $\sigma_{6}$
(Figs. \ref{fig:Early-stages-of}(c), \ref{fig:EnhancedInhom}), being
conceptually close to the analytical model, as well as nanovoids with
a diameter $d=0.6nm$ as present in silica and with a diameter increased
by a factor of 10 (Figs. \ref{fig:Early-stages-of}, \ref{fig:Plasma-density-2D}).
The results in Fig. \ref{fig:Early-stages-of} have been produced
with a spatial resolution of $\triangle x=0.5nm$. For the larger
simulation volumes in Figs. \ref{fig:Plasma-density-2D} and \ref{fig:EnhancedInhom},
a much coarser resolution of $\triangle x=5nm$ has been used.

We find, that all three inhomogeneity models reproduce the initial
predictions of the analytical model and lead to an almost exponential
growth of identical plasma structures into polarization direction
(Fig.  \ref{fig:Early-stages-of}(a)). Additionally we observe, that
the results do not differ significantly if a coarser discretization
is used. We conclude, that the final plasma structure is indeed invariant
concerning the seed. However, the irradiation time or intensity needed
to initiate the quasi-exponential growth depends on the nature and
strength of the inhomogeneity. Based on Fig. \ref{fig:Local-intensity-enhancement}(c)
we can say, that growth is initiated if a carrier density of approximately
half the resonant density is reached locally. For $\lambda=800nm$
and $n=1.45$ this corresponds to a value of $\frac{\rho_{Mie}}{2}\approx5\times10^{27}m^{-3}$.

Growth in polarization direction only slows down as the resulting
structure leaves the sub-wavelength domain. Now the plasma acts as
a nanoantenna whose maximum reflectivity would be reached at a size
of $\frac{\lambda}{2n}$. However, there is strong back-reflection
already at smaller sizes. Due to constructive interference with the
incident wave, further ionization is stimulated along the negative
propagation direction (Fig.  \ref{fig:Plasma-density-2D}(b)) and
a new structure is formed in front of the old one. At this point,
the lateral growth of the initial structure is inhibited, leading
to a finite size in polarization direction. 

During few optical cycles, a new structure is formed at the intensity
maximum caused by reflection from the previous one. In this way, a
periodic plasma chain with wavelength dimensions is initiated by a
tiny seed inhomogeneity of almost arbitrary nature (compare Figs.
 \ref{fig:Plasma-density-2D}(c) and \ref{fig:EnhancedInhom}) and
grows backwards against the propagation direction. 

\begin{figure}
\includegraphics[scale=0.55]{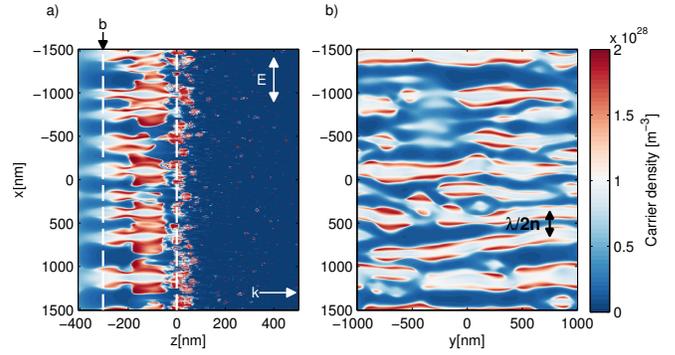}\caption{\label{fig:halfspace}(color online). Carrier density generated by
a plane wave (amplitude $E_{0}=1.7\times10^{10}Vm^{-1}$) normally
incident on a half space (z>0) filled with inhomogeneities($p_{inh}=0.01$,
$\sigma_{6,inh}=60\sigma_{6}$). Structures grow backward from the
inhomogeneous/homogeneous border at $z=0$ and form a grating with
a period of $\sim\frac{\lambda}{2n}=275nm$. Polarization and laser
propagation direction are indicated in panel a). Panel b) shows a
cut through the grating planes at $z=-300nm$.}
\end{figure}

\begin{figure}[t]
\includegraphics[scale=0.55]{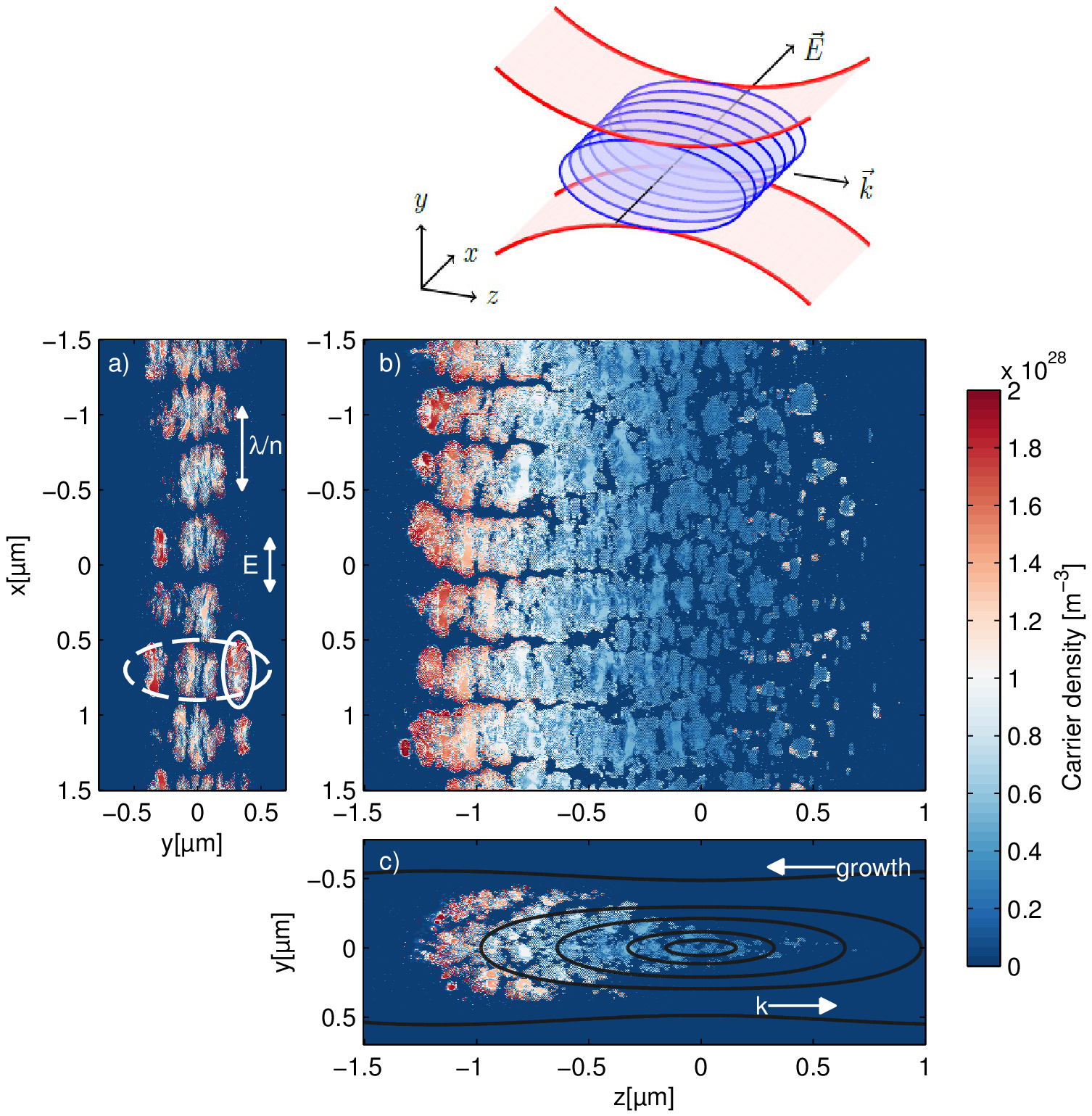}\caption{\label{fig:2dfocus3dsim}(color online). Carrier density within an
inhomogeneous volume ($p_{inh}=0.01$, $\sigma_{6,inh}=40\times\sigma_{6}$)
illuminated with a beam (maximum field strength in the homogeneous
case $E_{0}=1.9\times10^{10}Vm^{-1}$, $NA=0.8$) focussed in y-direction,
polarized in x-direction and propagating in z-direction. The linear
focus is located at $z=0$ (see overlay of lines of equal field strength
$|E|^{2}$ in panel c)). The sketch shows the focussing geometry and
orientation of periodic structures.}
\end{figure}

\begin{figure}
\includegraphics[scale=0.55]{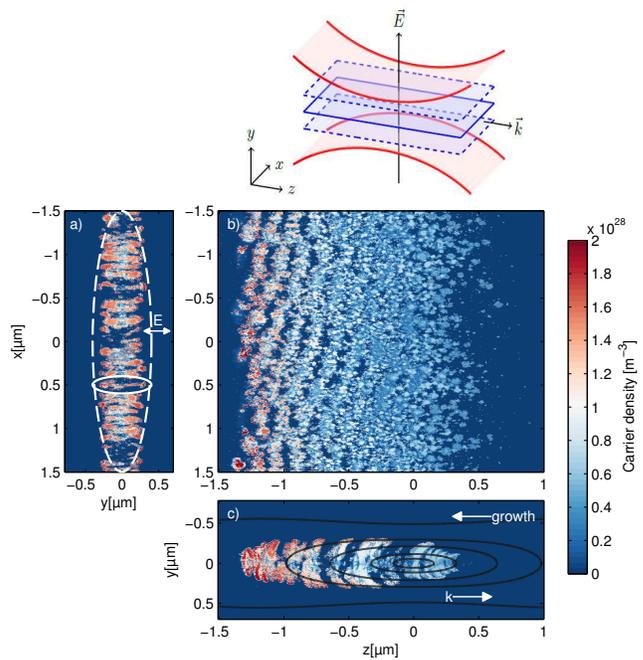}

\caption{\label{fig:2df3dsPolY}(color online). Carrier density within an inhomogeneous
volume ($p_{inh}=0.01$, $\sigma_{6,inh}=40\times\sigma_{6}$) illuminated
with a beam (maximum field strength in the homogeneous case $E_{0}=1.9\times10^{10}Vm^{-1}$,
$NA=0.8$) focussed and polarized in y-direction, and propagating
in z-direction. The linear focus is located at $z=0$ (see overlay
of lines of equal field strength $|E|^{2}$ in panel c)). The sketch
shows the focussing geometry and orientation of ionization structures.}
\end{figure}

\begin{figure}
\includegraphics[scale=0.55]{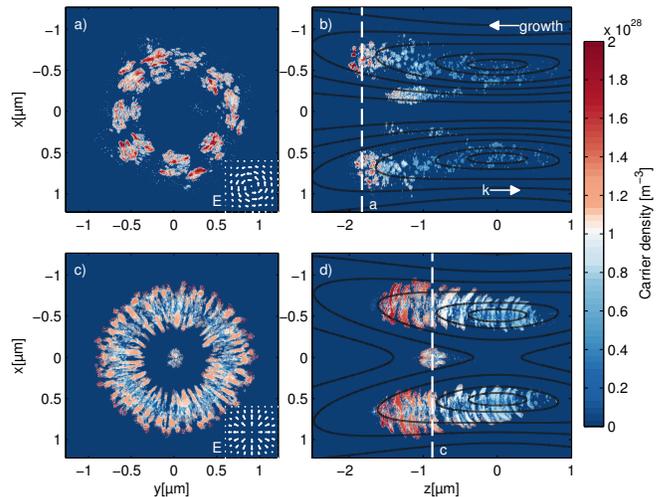}\caption{\label{fig:azi}(color online). Carrier density inside a volume filled
with inhomogeneities ($p_{inh}=0.01$) and irradiated with beams with
local polarization structure (maximum field strength in the homogeneous
case $E_{0}=1.7\times10^{10}Vm^{-1}$, $NA=0.5$) propagating in z-direction.
a) and b): Azimuthally polarized beam, $\sigma_{6,inh}=60\sigma_{6}$.
c) and d): Radially polarized beam, $\sigma_{6,inh}=30\sigma_{6}$.
Lines of equal field strength $|E|^{2}$ as expected in the linear
case are overlaid on panels b) and d).}
\end{figure}

We now turn to the study of randomly distributed inhomogeneities in
a three-dimensional volume. In all the simulations presented here,
we place pixel-sized inhomogeneities with a density $p_{inh}$ and
leave the background medium unperturbed. As expected we again observe
structure growth starting from the individual seed inhomogeneities.
For nanovoids or regions with a weakly enhanced ionization cross-section
as described above, the structures are sparsely distributed throughout
the material volume and do not interact. However for higher densities,
a large number of plasma stuctures competes in the growth process,
resulting in an onset of self-organization. To explore such effects,
we now allow for stronger ionization-enhancement inside the inhomogeneities
(Figs. \ref{fig:halfspace}-\ref{fig:azi}). 

In such systems, we observe the dense growth of the structures we
already described for the two-dimensional case. During their backwards
directed growth, the structures also expand into the third direction
not covered by our previous two-dimensional simulations. They do so
until they merge with their neighbours to form extended plasma planes
oriented perpendicularly to the polarization direction. Due to destructive
interference of scattered and incident light, ionization is suppressed
directly adjacent to each plasma plane and enhanced at a distance
of approximately $\frac{\lambda}{n}$. We note, that the field supression
and enhancement pattern in polarization direction around a single
complex plasma structure is still simliar to the one for a small plasma
sphere despite the differences in shape and size.

This effect leads to an interaction between separate structures. As
a result, order emerges during growth and a periodic pattern can be
formed. Since many structures form simultaneously and interaction
only becomes relevant during the growth process, the resulting period
is not completely determined by the position of the intensity maximum,
but also depends on growth conditions like the density and ionization
cross section of seed inhomogeneities or the intensity of the excitation. 

The smallest period can be observed under plane wave illumination.
In this case, structures closest to the source plane tend to grow
first and scatter light backwards, thus preventing any growth inside
the simulation volume behind them. To observe the free growth and
competition between different structures, we only fill a subspace
with inhomogeneities, leaving the space close to the source unperturbed.
Plasma structures form mainly at the border of the inhomogeneous region
and grow backwards into the unperturbed region, where intensity is
high (Fig. \ref{fig:halfspace}). Now only strong supression close
to the individual structures inhibits the growth of their neighbours,
resulting in a period as small as $\frac{\lambda}{2n}$. Note that
although growth starts at an interface, the resulting structure formation
remains a volume effect which can only to a certain extent be compared
to surface grating formation, which has been explained in terms of
interference of dipole radiation initiated at a rough material surface.\cite{PhysRevB.27.1141,PhysRevB.85.075320}
In our case, the self-organized period is caused by the growth of
densely arranged metallic plasma structures initiated at the inhomogeneities
on the interface. 

In more realistic simulations, we fill the entire simulation volume
with inhomogeneities and use focussed sources to control the location
of initial structure growth. First we use a source polarized in x-direction
and focussed only in y-direction, with z being the propagation direction
(see Fig. \ref{fig:2dfocus3dsim}). In x-direction the source profile
has an infinite size to allow for the formation of many grating planes.
In y-direction the beam is assumed to have a gaussian shape. As expected,
structures first emerge in the focal volume and grow backwards over
micrometer distances into regions of decreasing intensity (Fig. \ref{fig:2dfocus3dsim}).
Self-organization in this case is dominated by mutual enhancement
and we observe periods around $\frac{\lambda}{n}$. Since the planes
form inside an initially disordered volume, self-organization only
starts after a certain distance of backwards growth outside the original
focus. As can be seen in Fig. \ref{fig:2dfocus3dsim}(c), the transverse
size of the ionized region increases at the distance where self-organization
sets in (approximately $500nm$ behind the original focus), leading
to a ``carrot-shaped'' growth. We observe, that the individual grating
planes\textbf{,} as highlighted with a dashed ellipse in Fig. \ref{fig:2dfocus3dsim}(a)),
consist of several smaller structures, as highlighted with a second
ellipse, having a finite size below $\frac{\lambda}{2n}$ in polarization
direction. These correspond to the original wing-structure as it was
observed in Fig. \ref{fig:Plasma-density-2D}(a). Simliar to the case
of Fig. \ref{fig:halfspace}, the individual structures tend to merge
in the direction perpendicular to the polarization. However due to
the random distribution of seed inhomogeneities, the emerging structures
are not as regular as the ones observed previously.

To further demonstrate the effect of polarization, we continue using
the beam geometry of Fig. \ref{fig:2dfocus3dsim}, but choose an excitation
polarized in y-direction (Fig. \ref{fig:2df3dsPolY}). In this case,
one would expect the grating planes to be extended along the x- and
z- directions and arranged periodically along the y-direction. Since
the exciting beam has a limited size in y-direction, we only observe
a single structure, corresponding to a single grating plane (Fig.
\ref{fig:2df3dsPolY} (c)). As one would expect, no periodicity in
x-direction can be found (Fig. \ref{fig:2df3dsPolY}(a)). However,
similiar to Fig.\ref{fig:2dfocus3dsim}(a), the grating plane consists
of smaller structures, which have grown from seed inhomogeneities
and which have partly merged in x-direction. As suggested by the plasma
structures in Figs. \ref{fig:Plasma-density-2D}(c) and \ref{fig:EnhancedInhom},
we also observe some periodicity along the propagation direction (Fig.
\ref{fig:2df3dsPolY} (b)). 

We further consider spatially localized radially and azimuthally polarized
beams. Again we observe the formation of plasma planes perpendicular
to the local polarization. For an azimuthally polarized beam, this
leads to a star-shaped pattern containing several planes (Fig. \ref{fig:azi}(a),(b)).
In the radially polarized case, we obtain a single ring structure
caused by the transverse field components and a small structure in
the beam center caused by the maximum of the longitudinal component
(Fig. \ref{fig:azi}(c),(d)). Again, the individual planes in the
star-shaped pattern in Figs.\ref{fig:azi}(a),(b) and the single ring
structure in Figs.\ref{fig:azi}(c),(d) are made up of smaller structures
which tend to merge in the directions perpendicular to the exciting
polarization. In general the plasma planes generated by radially polarized
beams are more regular and cohesive. In that case, the polarization-enforced
ring-structure almost coincides with the region of maximum electric
field strength. In contrast, plasma planes generated by an azimuthallly
polarized beam point in radial direction, extending perpendicularly
to the region of maximum field strength. 

\begin{figure}
\includegraphics[scale=0.55]{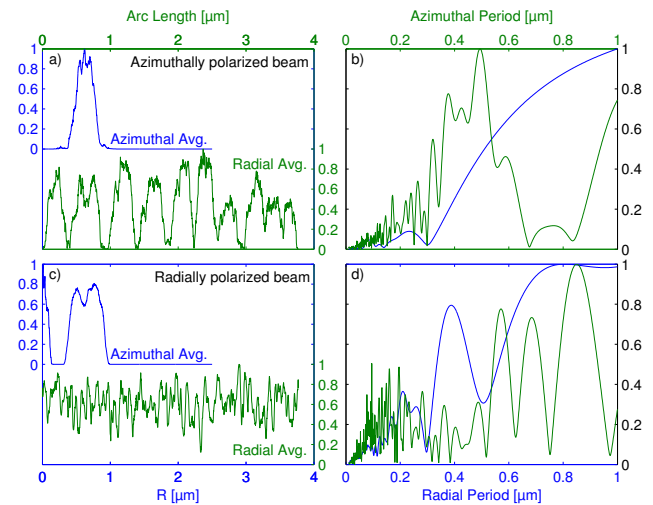}

\caption{\label{fig:FouRadiAzi}(color online). Structural analysis of the
data in Fig.\ref{fig:azi}. a): Directional averages in both azimuthal
and radial directions of the data in Fig. \ref{fig:azi}(a). b): Fourier
transforms of the data in a) plotted for structure periods below $1\mu m$.
c): Directional averages of the data in Fig. \ref{fig:azi}(c). d):
Fourier transforms of the data in c) plotted for structure periods
below $1\mu m$. To obtain the directional averages, the image data
has been transformed to a radial coordinate system and averaged over
each axis. The radial averages are plotted over the arc length $l=\theta R_{structure}$
for a nominal structure radius of $R_{structure}=600nm$.}
\end{figure}

To perform a more quantitative analysis of the ionization patterns
in Fig. \ref{fig:azi}, we transformed the data of Fig. \ref{fig:azi}(a)
and (c) into a radial coordinate system and took directional averages
in both radial and azimuthal directions (Fig. \ref{fig:FouRadiAzi}(a)
,(c)). For the azimuthally polarized beam, the radial averages exhibit
a strong periodic modulation in azimuthal direction. In the case of
the radially polarized beam, the modulation is weak and high carrier
densities above half the maximum are present across the whole ring
structure. The azimuthal averages show the finite size of the plasma
structure, which is determined by the finite beam size for both polarization
conditions. Figs. \ref{fig:FouRadiAzi}(b) and (d) show respective
fourier transforms. Only in the case of the azimuthally polarized
beam (Fig. \ref{fig:FouRadiAzi}(b)) a distinct periodicity of about
$500nm$ in azimuthal direction can be identified in the radial average
(green line).

\section{Relation to Experimental Results}

In the simulations of our model systems with randomly distributed
inhomogenities, we note strong similarities to experimentally observed
volume nanogratings both for beams with spatially homogeneous polarization\cite{springerlink:10.1007/s00339-011-6489-7}
and beams with a local polarization structure.\cite{Hnatovsky2011}

\textcolor{black}{Our physical model describes localized ionization
and the subsequent growth of nanoplasma, while neglecting the hydrodynamic
behaviour of the generated plasma. It thus corresponds to the ``nanoplasmonic''
model of grating formation suggested in publications by Hnatovski
et al.\cite{0953-4075-40-11-S03}. Contrary to the hypothesis presented
there, we do not observe any growth of extended plasma structures
for dielectric plasma densities. However close to and above the resonant
plasma density where $\varepsilon_{plasma}=-2\varepsilon_{background}$
we do observe the stimulated growth of ionized regions. Although the
structures initially grow into the polarization direction, they later
organize into thick planes with an orientation orthogonal to the incident
electric field.}

Despite these similarities with experimental observations our results
can only be regarded as a first attempt to understand the laser induced
formation of volume gratings in bulk glasses. Due to numerical complexity
of the subject we had to simplify the model considerably. Up to now
we did not model the evolution of the material in the excited region
after the pulse. In fact, real nanogratings form only over the course
of many laser pulses. Between individual pulses, the conduction band
carriers recombine completely, leaving only chemical and mechanical
material modifications as a feedback mechanism for further pulses.\cite{Richter:12,PhysRevLett.97.253001,Lancry:11}\emph{
}Self-organization emerges as a cumulative effect, and therefore quite
likely requires much lower ionization rates. 

In our model systems, inhomogeneities are strongly enhanced, allowing
for faster ionization of the material. Thus, the purely optical aspects
of self-organization can be studied in the course of a single irradiation
cycle. Due to its dependence on polarization and wavelength, nanograting
formation must be expected to be dominated by optical processes. Consequently,
our model should be able to capture some of the essential physics
involved.\emph{ }\textcolor{black}{Indeed},\emph{ }self-organization
of planes with the correct orientation and a wavelength-related period
is reproduced. In the case of an inhomogeneous half-space under planewave
irradiation, the commonly observed nanograting period of $\frac{\lambda}{2n}$
is matched by the simulated plasma structures. While most publications
focussing on high pulse numbers and fully formed gratings report on
thin planes with a period of $\frac{\lambda}{2n}$, also larger periods
around $\frac{\lambda}{n}$\cite{Mishchik2010} as well as much thicker
grating planes\cite{Taylor2008} have been observed in experiments
with low pulse numbers. Since our model does not include the material
modification between pulses, it is not surprising that we observe
similiar features in the majority of our simulations.

Our results further reproduce the large size as well as the increase in self-organization along the negative z-axis of nanogratings .\cite{Taylor2008} Both observations
can be explained by the backward growth of plasma structures, which
is driven by nearfield enhancement and continues well out of the focal
region. Experiments have shown, that a similiar growth takes place
over the course of multiple laser pulses during the generation of
nanogratings.\cite{Mishchik2010}

\section{Conclusion}

We have modelled the interaction of light with nanoscale inhomogeneities
in dielectrics undergoing multiphoton ionization. We observed, that
nanoscale inhomogeneities as e.g. voids influence the plasma formation
considerably. They induce the formation of large-scale plasma structures
with a final shape independent of the initial void. In case of randomly
distributed inhomogeneities we observe a strong interaction and subsequent
self-organization of evolving plasma structures reproducing some of
the key features of nanograting damage patterns in glass. 

Further research will include additional ionization mechanisms and
a more detailed description of the free carrier dynamics including
carrier heating, density and temperature dependent changes of the
collision frequency and hydrodynamic phenomena. To fully understand
the phenomenon of nanograting-formation, detailed simulations of the
material modifications taking place between pulses are required.
\begin{acknowledgments}
The authors gratefully acknowledge financial support by Deutsche Forschungsgemeinschaft
(priority program 1327: PE523/-2 and NO462/5-2) and the International
Max Planck Research School Physics of Light.
\end{acknowledgments}


\begin{thebibliography}{32}
\expandafter\ifx\csname natexlab\endcsname\relax\def\natexlab#1{#1}\fi
\expandafter\ifx\csname bibnamefont\endcsname\relax
  \def\bibnamefont#1{#1}\fi
\expandafter\ifx\csname bibfnamefont\endcsname\relax
  \def\bibfnamefont#1{#1}\fi
\expandafter\ifx\csname citenamefont\endcsname\relax
  \def\citenamefont#1{#1}\fi
\expandafter\ifx\csname url\endcsname\relax
  \def\url#1{\texttt{#1}}\fi
\expandafter\ifx\csname urlprefix\endcsname\relax\def\urlprefix{URL }\fi
\providecommand{\bibinfo}[2]{#2}
\providecommand{\eprint}[2][]{\url{#2}}

\bibitem[{\citenamefont{Mainfray and Manus}(1991)}]{0034-4885-54-10-002}
\bibinfo{author}{\bibfnamefont{G.}~\bibnamefont{Mainfray}} \bibnamefont{and}
  \bibinfo{author}{\bibfnamefont{G.}~\bibnamefont{Manus}},
  \bibinfo{journal}{Rep. Prog. Phys.} \textbf{\bibinfo{volume}{54}},
  \bibinfo{pages}{1333} (\bibinfo{year}{1991}),
  \urlprefix\url{http://stacks.iop.org/0034-4885/54/i=10/a=002}.

\bibitem[{\citenamefont{Davis et~al.}(1996)\citenamefont{Davis, Miura,
  Sugimoto, and Hirao}}]{Davis:96}
\bibinfo{author}{\bibfnamefont{K.~M.} \bibnamefont{Davis}},
  \bibinfo{author}{\bibfnamefont{K.}~\bibnamefont{Miura}},
  \bibinfo{author}{\bibfnamefont{N.}~\bibnamefont{Sugimoto}}, \bibnamefont{and}
  \bibinfo{author}{\bibfnamefont{K.}~\bibnamefont{Hirao}},
  \bibinfo{journal}{Opt. Lett.} \textbf{\bibinfo{volume}{21}},
  \bibinfo{pages}{1729} (\bibinfo{year}{1996}),
  \urlprefix\url{http://ol.osa.org/abstract.cfm?URI=ol-21-21-1729}.

\bibitem[{\citenamefont{Glezer and Mazur}(1997)}]{glezer:882}
\bibinfo{author}{\bibfnamefont{E.~N.} \bibnamefont{Glezer}} \bibnamefont{and}
  \bibinfo{author}{\bibfnamefont{E.}~\bibnamefont{Mazur}},
  \bibinfo{journal}{Appl. Phys. Lett.} \textbf{\bibinfo{volume}{71}},
  \bibinfo{pages}{882} (\bibinfo{year}{1997}),
  \urlprefix\url{http://link.aip.org/link/?APL/71/882/1}.

\bibitem[{\citenamefont{Shimotsuma et~al.}(2005)\citenamefont{Shimotsuma,
  Hirao, Qiu, and Kazansky}}]{eps20999}
\bibinfo{author}{\bibfnamefont{Y.}~\bibnamefont{Shimotsuma}},
  \bibinfo{author}{\bibfnamefont{K.}~\bibnamefont{Hirao}},
  \bibinfo{author}{\bibfnamefont{J.}~\bibnamefont{Qiu}}, \bibnamefont{and}
  \bibinfo{author}{\bibfnamefont{P.~G.} \bibnamefont{Kazansky}},
  \bibinfo{journal}{Mod. Phys. Lett. B} \textbf{\bibinfo{volume}{19}},
  \bibinfo{pages}{225} (\bibinfo{year}{2005}),
  \urlprefix\url{http://eprints.soton.ac.uk/20999/}.

\bibitem[{\citenamefont{Bhardwaj et~al.}(2006)\citenamefont{Bhardwaj, Simova,
  Rajeev, Hnatovsky, Taylor, Rayner, and Corkum}}]{PhysRevLett.96.057404}
\bibinfo{author}{\bibfnamefont{V.~R.} \bibnamefont{Bhardwaj}},
  \bibinfo{author}{\bibfnamefont{E.}~\bibnamefont{Simova}},
  \bibinfo{author}{\bibfnamefont{P.~P.} \bibnamefont{Rajeev}},
  \bibinfo{author}{\bibfnamefont{C.}~\bibnamefont{Hnatovsky}},
  \bibinfo{author}{\bibfnamefont{R.~S.} \bibnamefont{Taylor}},
  \bibinfo{author}{\bibfnamefont{D.~M.} \bibnamefont{Rayner}},
  \bibnamefont{and} \bibinfo{author}{\bibfnamefont{P.~B.}
  \bibnamefont{Corkum}}, \bibinfo{journal}{Phys. Rev. Lett.}
  \textbf{\bibinfo{volume}{96}}, \bibinfo{pages}{057404}
  (\bibinfo{year}{2006}),
  \urlprefix\url{http://link.aps.org/doi/10.1103/PhysRevLett.96.057404}.

\bibitem[{\citenamefont{Hnatovsky et~al.}(2011)\citenamefont{Hnatovsky,
  Shvedov, Krolikowski, and Rode}}]{Hnatovsky2011}
\bibinfo{author}{\bibfnamefont{C.}~\bibnamefont{Hnatovsky}},
  \bibinfo{author}{\bibfnamefont{V.}~\bibnamefont{Shvedov}},
  \bibinfo{author}{\bibfnamefont{W.}~\bibnamefont{Krolikowski}},
  \bibnamefont{and} \bibinfo{author}{\bibfnamefont{A.}~\bibnamefont{Rode}},
  \bibinfo{journal}{Phys. Rev. Lett.} \textbf{\bibinfo{volume}{106}},
  \bibinfo{pages}{123901} (\bibinfo{year}{2011}),
  \urlprefix\url{http://link.aps.org/doi/10.1103/PhysRevLett.106.123901}.

\bibitem[{\citenamefont{Petrov and Davis}(2008)}]{0953-4075-41-2-025601}
\bibinfo{author}{\bibfnamefont{G.~M.} \bibnamefont{Petrov}} \bibnamefont{and}
  \bibinfo{author}{\bibfnamefont{J.}~\bibnamefont{Davis}}, \bibinfo{journal}{J.
  Phys. B: At., Mol. Opt. Phys.} \textbf{\bibinfo{volume}{41}},
  \bibinfo{pages}{025601} (\bibinfo{year}{2008}),
  \urlprefix\url{http://stacks.iop.org/0953-4075/41/i=2/a=025601}.

\bibitem[{\citenamefont{Bourgeade et~al.}(2010)\citenamefont{Bourgeade, M�zel,
  and Saut}}]{Bourgeade2010}
\bibinfo{author}{\bibfnamefont{A.}~\bibnamefont{Bourgeade}},
  \bibinfo{author}{\bibfnamefont{C.}~\bibnamefont{M�zel}}, \bibnamefont{and}
  \bibinfo{author}{\bibfnamefont{O.}~\bibnamefont{Saut}}, \bibinfo{journal}{J.
  Sci. Comput.} \textbf{\bibinfo{volume}{44}}, \bibinfo{pages}{170}
  (\bibinfo{year}{2010}), ISSN \bibinfo{issn}{0885-7474},
  \urlprefix\url{http://dx.doi.org/10.1007/s10915-010-9375-0}.

\bibitem[{\citenamefont{Bulgakova et~al.}(2013)\citenamefont{Bulgakova, Zhukov,
  and Meshcheryakov}}]{raey}
\bibinfo{author}{\bibfnamefont{N.~M.} \bibnamefont{Bulgakova}},
  \bibinfo{author}{\bibfnamefont{V.~P.} \bibnamefont{Zhukov}},
  \bibnamefont{and} \bibinfo{author}{\bibfnamefont{Y.~P.}
  \bibnamefont{Meshcheryakov}}, \bibinfo{journal}{Appl. Phys. B} pp.
  \bibinfo{pages}{1--13} (\bibinfo{year}{2013}), ISSN
  \bibinfo{issn}{0946-2171},
  \urlprefix\url{http://dx.doi.org/10.1007/s00340-013-5488-0}.

\bibitem[{\citenamefont{Mezel et~al.}(2008)\citenamefont{Mezel, Hallo,
  Bourgeade, Hebert, Tikhonchuk, Chimier, Nkonga, Schurtz, and
  Travaille}}]{mezel:093504}
\bibinfo{author}{\bibfnamefont{C.}~\bibnamefont{Mezel}},
  \bibinfo{author}{\bibfnamefont{L.}~\bibnamefont{Hallo}},
  \bibinfo{author}{\bibfnamefont{A.}~\bibnamefont{Bourgeade}},
  \bibinfo{author}{\bibfnamefont{D.}~\bibnamefont{Hebert}},
  \bibinfo{author}{\bibfnamefont{V.~T.} \bibnamefont{Tikhonchuk}},
  \bibinfo{author}{\bibfnamefont{B.}~\bibnamefont{Chimier}},
  \bibinfo{author}{\bibfnamefont{B.}~\bibnamefont{Nkonga}},
  \bibinfo{author}{\bibfnamefont{G.}~\bibnamefont{Schurtz}}, \bibnamefont{and}
  \bibinfo{author}{\bibfnamefont{G.}~\bibnamefont{Travaille}},
  \bibinfo{journal}{Phys. Plasmas} \textbf{\bibinfo{volume}{15}},
  \bibinfo{eid}{093504} (pages~\bibinfo{numpages}{10}) (\bibinfo{year}{2008}),
  \urlprefix\url{http://link.aip.org/link/?PHP/15/093504/1}.

\bibitem[{\citenamefont{Doremus}(1966)}]{JACE:JACE461}
\bibinfo{author}{\bibfnamefont{R.~H.} \bibnamefont{Doremus}},
  \bibinfo{journal}{J. Am. Ceram. Soc.} \textbf{\bibinfo{volume}{49}},
  \bibinfo{pages}{461} (\bibinfo{year}{1966}), ISSN \bibinfo{issn}{1551-2916},
  \urlprefix\url{http://dx.doi.org/10.1111/j.1151-2916.1966.tb13299.x}.

\bibitem[{\citenamefont{Rajeev et~al.}(2007)\citenamefont{Rajeev, Gertsvolf,
  Hnatovsky, Simova, Taylor, Corkum, Rayner, and
  Bhardwaj}}]{0953-4075-40-11-S03}
\bibinfo{author}{\bibfnamefont{P.~P.} \bibnamefont{Rajeev}},
  \bibinfo{author}{\bibfnamefont{M.}~\bibnamefont{Gertsvolf}},
  \bibinfo{author}{\bibfnamefont{C.}~\bibnamefont{Hnatovsky}},
  \bibinfo{author}{\bibfnamefont{E.}~\bibnamefont{Simova}},
  \bibinfo{author}{\bibfnamefont{R.~S.} \bibnamefont{Taylor}},
  \bibinfo{author}{\bibfnamefont{P.~B.} \bibnamefont{Corkum}},
  \bibinfo{author}{\bibfnamefont{D.~M.} \bibnamefont{Rayner}},
  \bibnamefont{and} \bibinfo{author}{\bibfnamefont{V.~R.}
  \bibnamefont{Bhardwaj}}, \bibinfo{journal}{J. Phys. B: At., Mol. Opt. Phys.}
  \textbf{\bibinfo{volume}{40}}, \bibinfo{pages}{S273} (\bibinfo{year}{2007}),
  \urlprefix\url{http://stacks.iop.org/0953-4075/40/i=11/a=S03}.

\bibitem[{\citenamefont{Efimenko and Kim}(2011)}]{PhysRevE.84.036408}
\bibinfo{author}{\bibfnamefont{E.~S.} \bibnamefont{Efimenko}} \bibnamefont{and}
  \bibinfo{author}{\bibfnamefont{A.~V.} \bibnamefont{Kim}},
  \bibinfo{journal}{Phys. Rev. E} \textbf{\bibinfo{volume}{84}},
  \bibinfo{pages}{036408} (\bibinfo{year}{2011}),
  \urlprefix\url{http://link.aps.org/doi/10.1103/PhysRevE.84.036408}.

\bibitem[{\citenamefont{Efimenko et~al.}(2009)\citenamefont{Efimenko, Kim, and
  Quiroga-Teixeiro}}]{PhysRevLett.102.015002}
\bibinfo{author}{\bibfnamefont{E.~S.} \bibnamefont{Efimenko}},
  \bibinfo{author}{\bibfnamefont{A.~V.} \bibnamefont{Kim}}, \bibnamefont{and}
  \bibinfo{author}{\bibfnamefont{M.}~\bibnamefont{Quiroga-Teixeiro}},
  \bibinfo{journal}{Phys. Rev. Lett.} \textbf{\bibinfo{volume}{102}},
  \bibinfo{pages}{015002} (\bibinfo{year}{2009}),
  \urlprefix\url{http://link.aps.org/doi/10.1103/PhysRevLett.102.015002}.

\bibitem[{\citenamefont{Pe\~nano et~al.}(2005)\citenamefont{Pe\~nano, Sprangle,
  Hafizi, Manheimer, and Zigler}}]{Penano2005}
\bibinfo{author}{\bibfnamefont{J.~R.} \bibnamefont{Pe\~nano}},
  \bibinfo{author}{\bibfnamefont{P.}~\bibnamefont{Sprangle}},
  \bibinfo{author}{\bibfnamefont{B.}~\bibnamefont{Hafizi}},
  \bibinfo{author}{\bibfnamefont{W.}~\bibnamefont{Manheimer}},
  \bibnamefont{and} \bibinfo{author}{\bibfnamefont{A.}~\bibnamefont{Zigler}},
  \bibinfo{journal}{Phys. Rev. E} \textbf{\bibinfo{volume}{72}},
  \bibinfo{pages}{036412} (\bibinfo{year}{2005}),
  \urlprefix\url{http://link.aps.org/doi/10.1103/PhysRevE.72.036412}.

\bibitem[{\citenamefont{Yee}(1966)}]{Yee66numericalsolution}
\bibinfo{author}{\bibfnamefont{K.~S.} \bibnamefont{Yee}},
  \bibinfo{journal}{IEEE Trans. Antennas and Propagation} pp.
  \bibinfo{pages}{302--307} (\bibinfo{year}{1966}).

\bibitem[{\citenamefont{Taflove and Hagness}(2005)}]{taflove:2005}
\bibinfo{author}{\bibfnamefont{A.}~\bibnamefont{Taflove}} \bibnamefont{and}
  \bibinfo{author}{\bibfnamefont{S.~C.} \bibnamefont{Hagness}},
  \emph{\bibinfo{title}{{Computational Electrodynamics: The Finite-Difference
  Time-Domain Method, Third Edition}}} (\bibinfo{publisher}{Artech House},
  \bibinfo{year}{2005}), \bibinfo{edition}{3rd} ed., ISBN
  \bibinfo{isbn}{1580538320},
  \urlprefix\url{http://www.worldcat.org/isbn/1580538320}.

\bibitem[{\citenamefont{Gamaly et~al.}(2006)\citenamefont{Gamaly, Juodkazis,
  Nishimura, Misawa, Luther-Davies, Hallo, Nicolai, and
  Tikhonchuk}}]{PhysRevB.73.214101}
\bibinfo{author}{\bibfnamefont{E.~G.} \bibnamefont{Gamaly}},
  \bibinfo{author}{\bibfnamefont{S.}~\bibnamefont{Juodkazis}},
  \bibinfo{author}{\bibfnamefont{K.}~\bibnamefont{Nishimura}},
  \bibinfo{author}{\bibfnamefont{H.}~\bibnamefont{Misawa}},
  \bibinfo{author}{\bibfnamefont{B.}~\bibnamefont{Luther-Davies}},
  \bibinfo{author}{\bibfnamefont{L.}~\bibnamefont{Hallo}},
  \bibinfo{author}{\bibfnamefont{P.}~\bibnamefont{Nicolai}}, \bibnamefont{and}
  \bibinfo{author}{\bibfnamefont{V.~T.} \bibnamefont{Tikhonchuk}},
  \bibinfo{journal}{Phys. Rev. B} \textbf{\bibinfo{volume}{73}},
  \bibinfo{pages}{214101} (\bibinfo{year}{2006}),
  \urlprefix\url{http://link.aps.org/doi/10.1103/PhysRevB.73.214101}.

\bibitem[{\citenamefont{Sudrie et~al.}(2002)\citenamefont{Sudrie, Couairon,
  Franco, Lamouroux, Prade, Tzortzakis, and
  Mysyrowicz}}]{PhysRevLett.89.186601}
\bibinfo{author}{\bibfnamefont{L.}~\bibnamefont{Sudrie}},
  \bibinfo{author}{\bibfnamefont{A.}~\bibnamefont{Couairon}},
  \bibinfo{author}{\bibfnamefont{M.}~\bibnamefont{Franco}},
  \bibinfo{author}{\bibfnamefont{B.}~\bibnamefont{Lamouroux}},
  \bibinfo{author}{\bibfnamefont{B.}~\bibnamefont{Prade}},
  \bibinfo{author}{\bibfnamefont{S.}~\bibnamefont{Tzortzakis}},
  \bibnamefont{and}
  \bibinfo{author}{\bibfnamefont{A.}~\bibnamefont{Mysyrowicz}},
  \bibinfo{journal}{Phys. Rev. Lett.} \textbf{\bibinfo{volume}{89}},
  \bibinfo{pages}{186601} (\bibinfo{year}{2002}),
  \urlprefix\url{http://link.aps.org/doi/10.1103/PhysRevLett.89.186601}.

\bibitem[{\citenamefont{Mao et~al.}(2003)\citenamefont{Mao, Mao, and
  Russo}}]{mao:697}
\bibinfo{author}{\bibfnamefont{X.}~\bibnamefont{Mao}},
  \bibinfo{author}{\bibfnamefont{S.~S.} \bibnamefont{Mao}}, \bibnamefont{and}
  \bibinfo{author}{\bibfnamefont{R.~E.} \bibnamefont{Russo}},
  \bibinfo{journal}{Appl. Phys. Lett.} \textbf{\bibinfo{volume}{82}},
  \bibinfo{pages}{697} (\bibinfo{year}{2003}),
  \urlprefix\url{http://link.aip.org/link/?APL/82/697/1}.

\bibitem[{\citenamefont{Audebert et~al.}(1994)\citenamefont{Audebert, Daguzan,
  Dos~Santos, Gauthier, Geindre, Guizard, Hamoniaux, Krastev, Martin, Petite
  et~al.}}]{PhysRevLett.73.1990}
\bibinfo{author}{\bibfnamefont{P.}~\bibnamefont{Audebert}},
  \bibinfo{author}{\bibfnamefont{P.}~\bibnamefont{Daguzan}},
  \bibinfo{author}{\bibfnamefont{A.}~\bibnamefont{Dos~Santos}},
  \bibinfo{author}{\bibfnamefont{J.~C.} \bibnamefont{Gauthier}},
  \bibinfo{author}{\bibfnamefont{J.~P.} \bibnamefont{Geindre}},
  \bibinfo{author}{\bibfnamefont{S.}~\bibnamefont{Guizard}},
  \bibinfo{author}{\bibfnamefont{G.}~\bibnamefont{Hamoniaux}},
  \bibinfo{author}{\bibfnamefont{K.}~\bibnamefont{Krastev}},
  \bibinfo{author}{\bibfnamefont{P.}~\bibnamefont{Martin}},
  \bibinfo{author}{\bibfnamefont{G.}~\bibnamefont{Petite}},
  \bibnamefont{et~al.}, \bibinfo{journal}{Phys. Rev. Lett.}
  \textbf{\bibinfo{volume}{73}}, \bibinfo{pages}{1990} (\bibinfo{year}{1994}),
  \urlprefix\url{http://link.aps.org/doi/10.1103/PhysRevLett.73.1990}.

\bibitem[{\citenamefont{Richter et~al.}(2012)\citenamefont{Richter, Jia,
  Heinrich, D\"{o}ring, Peschel, T\"{u}nnermann, and Nolte}}]{Richter:12}
\bibinfo{author}{\bibfnamefont{S.}~\bibnamefont{Richter}},
  \bibinfo{author}{\bibfnamefont{F.}~\bibnamefont{Jia}},
  \bibinfo{author}{\bibfnamefont{M.}~\bibnamefont{Heinrich}},
  \bibinfo{author}{\bibfnamefont{S.}~\bibnamefont{D\"{o}ring}},
  \bibinfo{author}{\bibfnamefont{U.}~\bibnamefont{Peschel}},
  \bibinfo{author}{\bibfnamefont{A.}~\bibnamefont{T\"{u}nnermann}},
  \bibnamefont{and} \bibinfo{author}{\bibfnamefont{S.}~\bibnamefont{Nolte}},
  \bibinfo{journal}{Opt. Lett.} \textbf{\bibinfo{volume}{37}},
  \bibinfo{pages}{482} (\bibinfo{year}{2012}),
  \urlprefix\url{http://ol.osa.org/abstract.cfm?URI=ol-37-4-482}.

\bibitem[{\citenamefont{Hasegawa et~al.}(2000)\citenamefont{Hasegawa, Saneyasu,
  Tabata, Tang, Nagai, Chiba, and Ito}}]{Hasegawa2000}
\bibinfo{author}{\bibfnamefont{M.}~\bibnamefont{Hasegawa}},
  \bibinfo{author}{\bibfnamefont{M.}~\bibnamefont{Saneyasu}},
  \bibinfo{author}{\bibfnamefont{M.}~\bibnamefont{Tabata}},
  \bibinfo{author}{\bibfnamefont{Z.}~\bibnamefont{Tang}},
  \bibinfo{author}{\bibfnamefont{Y.}~\bibnamefont{Nagai}},
  \bibinfo{author}{\bibfnamefont{T.}~\bibnamefont{Chiba}}, \bibnamefont{and}
  \bibinfo{author}{\bibfnamefont{Y.}~\bibnamefont{Ito}},
  \bibinfo{journal}{Nuclear Instruments and Methods in Physics Research Section
  B: Beam Interactions with Materials and Atoms}
  \textbf{\bibinfo{volume}{166-167}}, \bibinfo{pages}{431}
  (\bibinfo{year}{2000}), ISSN \bibinfo{issn}{0168583X},
  \urlprefix\url{http://linkinghub.elsevier.com/retrieve/pii/S0168583X99010265}.

\bibitem[{\citenamefont{Rajeev et~al.}(2006)\citenamefont{Rajeev, Gertsvolf,
  Simova, Hnatovsky, Taylor, Bhardwaj, Rayner, and
  Corkum}}]{PhysRevLett.97.253001}
\bibinfo{author}{\bibfnamefont{P.~P.} \bibnamefont{Rajeev}},
  \bibinfo{author}{\bibfnamefont{M.}~\bibnamefont{Gertsvolf}},
  \bibinfo{author}{\bibfnamefont{E.}~\bibnamefont{Simova}},
  \bibinfo{author}{\bibfnamefont{C.}~\bibnamefont{Hnatovsky}},
  \bibinfo{author}{\bibfnamefont{R.~S.} \bibnamefont{Taylor}},
  \bibinfo{author}{\bibfnamefont{V.~R.} \bibnamefont{Bhardwaj}},
  \bibinfo{author}{\bibfnamefont{D.~M.} \bibnamefont{Rayner}},
  \bibnamefont{and} \bibinfo{author}{\bibfnamefont{P.~B.}
  \bibnamefont{Corkum}}, \bibinfo{journal}{Phys. Rev. Lett.}
  \textbf{\bibinfo{volume}{97}}, \bibinfo{pages}{253001}
  (\bibinfo{year}{2006}),
  \urlprefix\url{http://link.aps.org/doi/10.1103/PhysRevLett.97.253001}.

\bibitem[{\citenamefont{Lancry et~al.}(2011)\citenamefont{Lancry, Poumellec,
  Cook, and Canning}}]{Lancry:11}
\bibinfo{author}{\bibfnamefont{M.}~\bibnamefont{Lancry}},
  \bibinfo{author}{\bibfnamefont{B.}~\bibnamefont{Poumellec}},
  \bibinfo{author}{\bibfnamefont{K.}~\bibnamefont{Cook}}, \bibnamefont{and}
  \bibinfo{author}{\bibfnamefont{J.}~\bibnamefont{Canning}}, in
  \emph{\bibinfo{booktitle}{Proceedings of the International Quantum
  Electronics Conference and Conference on Lasers and Electro-Optics Pacific
  Rim 2011}} (\bibinfo{publisher}{Optical Society of America},
  \bibinfo{year}{2011}), p. \bibinfo{pages}{C229},
  \urlprefix\url{http://www.opticsinfobase.org/abstract.cfm?URI=CLEOPR-2011-C229}.

\bibitem[{\citenamefont{Bohren and Huffman}(1983)}]{Bohren_Huffman_1983}
\bibinfo{author}{\bibfnamefont{C.~F.} \bibnamefont{Bohren}} \bibnamefont{and}
  \bibinfo{author}{\bibfnamefont{D.~R.} \bibnamefont{Huffman}},
  \emph{\bibinfo{title}{Absorption and scattering of light by small
  particles}}, vol.~\bibinfo{volume}{1} (\bibinfo{publisher}{Wiley},
  \bibinfo{year}{1983}),
  \urlprefix\url{http://adsabs.harvard.edu/abs/1983uaz..rept.....B}.

\bibitem[{\citenamefont{Jackson}(1999)}]{jackson_classical_1999}
\bibinfo{author}{\bibfnamefont{J.~D.} \bibnamefont{Jackson}},
  \emph{\bibinfo{title}{Classical electrodynamics}}
  (\bibinfo{publisher}{Wiley}, \bibinfo{address}{New York, {NY}},
  \bibinfo{year}{1999}), \bibinfo{edition}{3rd} ed., ISBN
  \bibinfo{isbn}{9780471309321},
  \urlprefix\url{http://cdsweb.cern.ch/record/490457}.

\bibitem[{\citenamefont{Sipe et~al.}(1983)\citenamefont{Sipe, Young, Preston,
  and van Driel}}]{PhysRevB.27.1141}
\bibinfo{author}{\bibfnamefont{J.~E.} \bibnamefont{Sipe}},
  \bibinfo{author}{\bibfnamefont{J.~F.} \bibnamefont{Young}},
  \bibinfo{author}{\bibfnamefont{J.~S.} \bibnamefont{Preston}},
  \bibnamefont{and} \bibinfo{author}{\bibfnamefont{H.~M.} \bibnamefont{van
  Driel}}, \bibinfo{journal}{Phys. Rev. B} \textbf{\bibinfo{volume}{27}},
  \bibinfo{pages}{1141} (\bibinfo{year}{1983}),
  \urlprefix\url{http://link.aps.org/doi/10.1103/PhysRevB.27.1141}.

\bibitem[{\citenamefont{Skolski et~al.}(2012)\citenamefont{Skolski, R\"omer,
  Obona, Ocelik, Huis in~'t Veld, and De~Hosson}}]{PhysRevB.85.075320}
\bibinfo{author}{\bibfnamefont{J.~Z.~P.} \bibnamefont{Skolski}},
  \bibinfo{author}{\bibfnamefont{G.~R. B.~E.} \bibnamefont{R\"omer}},
  \bibinfo{author}{\bibfnamefont{J.~V.} \bibnamefont{Obona}},
  \bibinfo{author}{\bibfnamefont{V.}~\bibnamefont{Ocelik}},
  \bibinfo{author}{\bibfnamefont{A.~J.} \bibnamefont{Huis in~'t Veld}},
  \bibnamefont{and} \bibinfo{author}{\bibfnamefont{J.~T.~M.}
  \bibnamefont{De~Hosson}}, \bibinfo{journal}{Phys. Rev. B}
  \textbf{\bibinfo{volume}{85}}, \bibinfo{pages}{075320}
  (\bibinfo{year}{2012}),
  \urlprefix\url{http://link.aps.org/doi/10.1103/PhysRevB.85.075320}.

\bibitem[{\citenamefont{Richter et~al.}(2011)\citenamefont{Richter, Heinrich,
  D�ring, T�nnermann, and Nolte}}]{springerlink:10.1007/s00339-011-6489-7}
\bibinfo{author}{\bibfnamefont{S.}~\bibnamefont{Richter}},
  \bibinfo{author}{\bibfnamefont{M.}~\bibnamefont{Heinrich}},
  \bibinfo{author}{\bibfnamefont{S.}~\bibnamefont{D�ring}},
  \bibinfo{author}{\bibfnamefont{A.}~\bibnamefont{T�nnermann}},
  \bibnamefont{and} \bibinfo{author}{\bibfnamefont{S.}~\bibnamefont{Nolte}},
  \bibinfo{journal}{Appl. Phys. A: Mater. Sci. Process.}
  \textbf{\bibinfo{volume}{104}}, \bibinfo{pages}{503} (\bibinfo{year}{2011}),
  ISSN \bibinfo{issn}{0947-8396}, \bibinfo{note}{10.1007/s00339-011-6489-7},
  \urlprefix\url{http://dx.doi.org/10.1007/s00339-011-6489-7}.

\bibitem[{\citenamefont{Mishchik et~al.}(2010)\citenamefont{Mishchik, Cheng,
  Huo, Burakov, Mauclair, Mermillod-Blondin, Rosenfeld, Ouerdane, Boukenter,
  Parriaux et~al.}}]{Mishchik2010}
\bibinfo{author}{\bibfnamefont{K.}~\bibnamefont{Mishchik}},
  \bibinfo{author}{\bibfnamefont{G.}~\bibnamefont{Cheng}},
  \bibinfo{author}{\bibfnamefont{G.}~\bibnamefont{Huo}},
  \bibinfo{author}{\bibfnamefont{I.~M.} \bibnamefont{Burakov}},
  \bibinfo{author}{\bibfnamefont{C.}~\bibnamefont{Mauclair}},
  \bibinfo{author}{\bibfnamefont{a.}~\bibnamefont{Mermillod-Blondin}},
  \bibinfo{author}{\bibfnamefont{a.}~\bibnamefont{Rosenfeld}},
  \bibinfo{author}{\bibfnamefont{Y.}~\bibnamefont{Ouerdane}},
  \bibinfo{author}{\bibfnamefont{a.}~\bibnamefont{Boukenter}},
  \bibinfo{author}{\bibfnamefont{O.}~\bibnamefont{Parriaux}},
  \bibnamefont{et~al.}, \bibinfo{journal}{Optics express}
  \textbf{\bibinfo{volume}{18}}, \bibinfo{pages}{24809} (\bibinfo{year}{2010}),
  ISSN \bibinfo{issn}{1094-4087},
  \urlprefix\url{http://www.ncbi.nlm.nih.gov/pubmed/21164827}.

\bibitem[{\citenamefont{Taylor et~al.}(2008)\citenamefont{Taylor, Hnatovsky,
  and Simova}}]{Taylor2008}
\bibinfo{author}{\bibfnamefont{R.}~\bibnamefont{Taylor}},
  \bibinfo{author}{\bibfnamefont{C.}~\bibnamefont{Hnatovsky}},
  \bibnamefont{and} \bibinfo{author}{\bibfnamefont{E.}~\bibnamefont{Simova}},
  \bibinfo{journal}{Laser \& Photonics Review} \textbf{\bibinfo{volume}{2}},
  \bibinfo{pages}{26} (\bibinfo{year}{2008}), ISSN \bibinfo{issn}{18638880},
  \urlprefix\url{http://doi.wiley.com/10.1002/lpor.200710031}.

\end{thebibliography}
\end{document}